\documentstyle[aps,prb,epsf,multicol]{revtex}
\newcommand{\be}{\begin{equation}}
\newcommand{\ee}{\end{equation}}
\newcommand{\bea}{\begin{eqnarray}}
\newcommand{\eea}{\end{eqnarray}}

\begin{document}
\draft

\title{
Quasiparticle spectrum of a type-II superconductor in a high
magnetic field with randomly pinned vortices
}
\author{P.D. Sacramento} 
\address{
Departamento de F\'{\i}sica and CFIF, Instituto Superior T\'ecnico 
, Av. Rovisco Pais, P-1096 Lisboa Codex, Portugal}
\address{~
\parbox{14cm}{\rm
\medskip
We show that gapless superconductivity of a strongly type-II superconductor
in a high magnetic field prevails in the presence of disorder, suggesting a
topological nature. We calculate the density of states of the Bogoliubov-de Gennes
quasi-particles for a two-dimensional inhomogeneous system
in both cases of weak and strong disorder. 
In the limit of
very weak disorder, the effect is very small and the density of states is not
appreciably changed. As the disorder increases, the density of states at low energies increases
and the ratio of the low energy density of states to its maximum
increases significantly.}}

\maketitle
\begin{multicols}{2}
\narrowtext
The interplay between superconductivity and a magnetic field has attracted
interest for a long time. For sufficiently strong fields the Meissner phase
is destroyed and a mixed state appears in the form of a quantized vortex
lattice \cite{1}. The superconductor order parameter has zeros at the vortex
locations, through which the external magnetic field penetrates in the sample.
Contrarily to previous understanding, the increase of the magnetic field
intensity, and its associated diamagnetic pair breaking, is counteracted at
high magnetic fields by the Landau level structure of the electrons \cite{2}.
This leads to interesting properties such as enhancement of the superconducting
transition temperature at very high magnetic fields where the electrons are
confined to the lowest Landau level \cite{2}. Associated with the zeros of the
order parameter in real space are gapless points in the magnetic Brillouin
zone \cite{3}, which lead to qualitatively different behavior at low temperatures
and high magnetic fields \cite{4}.

It has been argued that the gapless behavior is restricted to high fields very
close to the upper critical line where the so-called diagonal approximation
(where the coupling between Landau levels is neglected) is valid. It has been
shown, however, that the presence of off-diagonal terms does not destroy this
behavior \cite{4} and that a perturbation scheme on the off-diagonal terms is
possible, as long as there are no band-crossings \cite{5}.
It was shown analytically to all orders in the perturbation theory on the
off-diagonal terms that there is always a discrete set of points which are
gapless that are associated with coherent propagation of the quasiparticles
(so-called Eilenberger points). These nodes are associated with the center
of mass coordinates of the Cooper pairs and not to some internal structure
like in d-wave superconductors. Lowering the magnetic field, a quantum level-crossing
transition has been found that eventually leads to a gapped regime and to
states localized in the vortex cores \cite{6,7}.

On the other hand, the effect of disorder on superconductivity has also attracted
interest for a long time. In the case of non-magnetic impurities and s-wave
pairing Anderson's theorem states that, at least for low concentrations, they
have little effect since the impurities are not pair-breaking \cite{8}. In d-wave
superconductors however, non-magnetic impurities cause a strong pair breaking
effect \cite{9}. In the limit of strong scattering it was found that the lowest energy
quasiparticles become localized below the mobility gap, even in a regime where
the single-electron wave-functions are still extended \cite{10}. This result has been
confirmed solving the Bogoliubov-de Gennes equations with a finite concentration
of non-magnetic impurities \cite{11}. However, allowing for angular dependent impurity
scattering potentials it has been found that the scattering processes close to
the gap nodes may give rise to extended gapless regions \cite{12}. The case of
magnetic impurities in the s-wave case also leads to gapless superconductivity \cite{13}.

The question we wish to address in this paper is if the presence of disorder
affects the gapless behavior found in the low-$T$ high magnetic field regime 
of s-wave superconductors discussed above. The case of a dirty but homogeneous
superconductor was considered before \cite{14}. It was assumed that the order parameter
is not significantly affected by the impurities and retains its periodic structure.
It was found that when the disorder becomes stronger than some critical value,
a finite density of states appears at the Fermi surface.

In general, since the interactions between the vortices are repulsive, 
a lattice structure
is more favorable energetically. However, if pinning centers are
present in the system, the higher energies of different configurations of the
vortices may be offset by the presence of disorder. The question then arises if
the gapless behavior prevails in this more general case. If the magnetic field
is very high, such that the system is in the quantum limit where the electrons
are confined to the lowest Landau level, it has been shown that for an arbitrary
configuration of zeros there is at least one gapless point \cite{15}. In this paper 
we will
consider a more general case in which the off-diagonal terms are included.

In the mean-field approximation the Hamiltonian of the superconducting system can
be diagonalized and the energy eigenvalues are the solutions of the Bogoliubov-de
Gennes equations (BdG) \cite{16}
\bea
\left[ \frac{1}{2m} \left( \vec{p}- \frac{e}{c} \vec{A} \right)^2 - \mu \right] u(
\vec{r}) + \Delta(\vec{r}) v(\vec{r}) & = & E u(\vec{r}) \nonumber \\
-\left[ \frac{1}{2m} \left( \vec{p}+ \frac{e}{c} \vec{A} \right)^2 - \mu \right] v(
\vec{r}) + \Delta^{*}(\vec{r}) u(\vec{r}) & = & E v(\vec{r})
\eea
where $\Delta(\vec{r})$ is the order parameter and $E$ the energy. If $\Delta(
\vec{r})=0$, the solutions are the Landau eigenfunctions which, for a two-dimensional
system perpendicular to the magnetic field, read in the Landau gauge
\be
\phi_{nq} = \frac{1}{\sqrt{L_x}} \frac{1}{\sqrt{l \sqrt{\pi} 2^n n!}} e^{iqx}
e^{-\frac{1}{2} \left[ \frac{y}{l} + ql \right]^2 } H_n \left[\frac{y}{l}+ql \right]
\ee
where $n$ is the Landau index, $L_x$ is the length of the system in the $x$
direction, $l$ is the magnetic length given by
$l^2=\hbar c/eH$ and $H_n$ is an Hermite polynomial. The
energy eigenvalues are those of an harmonic oscillator centered at $y_0=-ql^2$
\be
E_n=\hbar \omega_c \left( n + \frac{1}{2} \right)
\ee
where $\omega_c$ is the cyclotron frequency. Taking $L_y$ to be the dimension
along $y$ we obtain that $-L_y/2 \leq ql^2 \leq L_y/2$. In the presence of
$\Delta(\vec{r}) \neq 0$ we can use the Landau basis like
\bea
u(\vec{r}) & = & \sum_{nq} u_{nq} \phi_{nq}(\vec{r}) \nonumber \\
v(\vec{r}) & = & \sum_{nq} v_{nq} \phi_{nq}^{*}(\vec{r})
\eea
and obtain the corresponding eigensystem
\bea
\left[ n-n_c \right] u_{nk}^{\mu} + \sum_{mq}  
v_{mq}^{\mu} \Delta_{nm}^{kq} & = & \epsilon^{\mu} u_{nk}^{\mu}  \nonumber \\
-\left[ n-n_c \right] v_{nk}^{\mu} + \sum_{mq}  
u_{mq}^{\mu} \left(\Delta_{mn}^{qk} \right)^{*} & = & \epsilon^{\mu} v_{nk}^{\mu} 
\eea
where $\epsilon^{\mu} = E^{\mu}/(\hbar \omega_c)$, $n_c$ is defined by $\mu=\hbar
\omega_c (n_c + \frac{1}{2})$ and
\be
\Delta_{nm}^{kq} = \int d\vec{r} \phi_{nk}^{*} (\vec{r}) \frac{\Delta(\vec{r})}{
\hbar \omega_c} \phi_{mq}^{*} (\vec{r}).
\ee
The excitation spectrum is then obtained solving eqs. (5) with an appropriate choice for
the order parameter.

In the lattice case, Abrikosov's solution can be written in the Landau gauge as \cite{1}
\be
\Delta (\vec{r}) = \Delta \sum_{p} e^{i\pi \frac{b_x}{a} p^2} e^{i \frac{2\pi p}{a} x}
e^{-\left( \frac{y}{l} + \frac{\pi p}{a} l \right)^2}
\ee
The vortex lattice is characterized by unit vectors $\vec{a}=(a,0)$ and $\vec{b}=(b_x,b_y)$
where $b_x=0$, $b_y=a$ for a square lattice and $b_x=\frac{1}{2}a$, $b_y=\frac{\sqrt{3}}{2}a$ for
a triangular lattice. This form for the order parameter is valid sufficiently close to the
upper critical field, since it is entirely contained in the lowest Landau level of Cooper
charge $2e$. We will not consider contributions to the order parameter from higher Landau levels.
In this case the self-consistent equation for the order parameter reduces to a single relation
between $\Delta$ and $V$, the attractive interaction strength between the electrons.
In the following we will consider the square lattice ($L_x=L_y=L$), for simplicity. 
In this case the lattice
constant $a=l \sqrt{\pi}$ and the zeros are located at the points $x_i=(i+\frac{1}{2})l\sqrt{\pi}$,
 $y_j=(j+\frac{1}{2})l \sqrt{\pi}$.

An expression for the order parameter has also been found for an arbitrary 
distribution of the zeros
which in the symmetric gauge can be written as \cite{17}
\be
\Delta(x,y)=\bar{\Delta} \prod_{i=1}^{N_{\phi}} \left[ \frac{x-x_i}{l} + i \frac{y-y_i}{l} \right]
e^{-\frac{1}{2l^2 N_{\phi}} \left[ (x-x_i)^2 + (y-y_i)^2 \right] }
\ee
Here, $N_{\phi}$ is the number of vortices in the system (number of zeros $(x_i,y_i)$). The solution
of the BdG equations is then simply obtained numerically using eq. (6) and performing the gauge
transformation of eq. (8) to the Landau gauge.

In the lattice case it is more convenient to use a representation in terms of the magnetic
wave-functions \cite{18,4} instead of eq. (4), to take advantage of the 
translational invariance. The generalization
to a random configuration is however more conveniently done using a real space representation. We consider
a finite system and we have to take account of finite size effects. In particular, the use of eq. (8)
for $\Delta(\vec{r})$ is very sensitive. We have therefore compared eqs. (7) and (8) for a finite
system. (It is convenient to define new variables $X=\frac{x}{L}$, $Y=\frac{y}{L}$ and to use that
$L=l\sqrt{\pi N_{\phi}}$). The effect of the finite size was eliminated (less than $1\%$ 
difference with respect
to Abrikosov's solution) using periodic boundary conditions and calculating $\Delta (x,y)$ using
eq. (8) over a set of zeros contained in a circle of unit radius (recall that $-\frac{1}{2} \leq X,Y
\leq \frac{1}{2}$) and relating $\bar{\Delta}$ with $\Delta$ to 
give the same amplitude. Having obtained 
the excitation spectrum and the eigenvectors we calculated the density of states using the expression
\be
\rho(\omega) = \sum_{\mu} \left[ \sum_{nq} |u_{nq}^{\mu}|^2 \delta(\omega-\epsilon^{\mu}) 
+ \sum_{nq} |v_{nq}^{\mu}|^2 \delta(\omega+\epsilon^{\mu}) \right]
\ee
where the sum is restricted to $\epsilon^{\mu} \geq 0$.

We considered three cases: i) square lattice, ii) weakly disordered lattice and iii) randomly
pinned configuration (strong disorder). In the case of weak disorder we considered a distribution
of zeros $X_i=X_i^L + \delta (-\frac{1}{2}+r)/\sqrt{N_{\phi}}$, where $X_i^L$ is the regular lattice
location, $\delta$ is an adjustable amplitude and $r$ is a random number $0<r<1$ (and similarly for
$Y_i$). In the strong disorder case we allow $X_i,Y_i$ to take any values on the system. We calculated
the average over disorder directly on the density of states. That
is, we select randomly one configuration
of the zeros and 
\begin{figure}[h]
\epsfxsize=3.4in
\epsfysize=3.4in
\epsffile{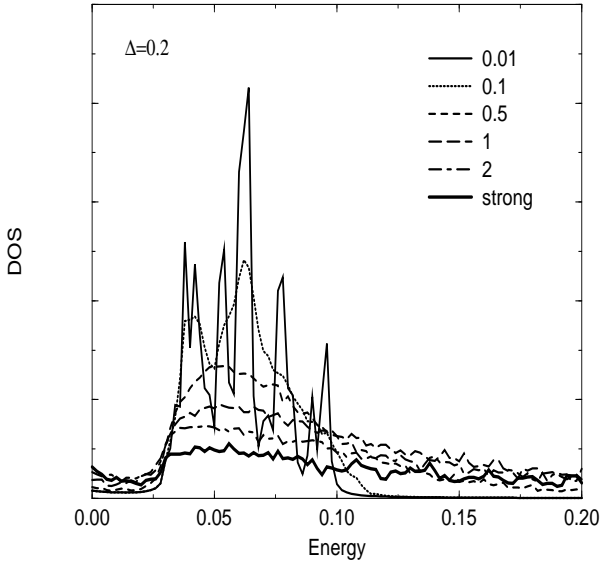}
\caption{
Density of states (DOS) (in arbitray units) as a function of energy 
for $n_c=10 \pm 2$, for
$\Delta=0.2$ for the cases of $\delta=0.01,0.1,0.5,1,2$ and for the case of strong
disorder with adjusted chemical potential (see text). Only the lowest band is shown.
}
\label{fig1}
\end{figure} 
\noindent obtain the excitation spectrum and $\rho(\omega)$. We repeat this process many
times and then we take the average over the resulting expressions for the density of states. This is
the final result.

To check the accuracy of the method we started with the lattice case 
previously solved \cite{4,5,6}. The
Cooper pairs are formed from electrons in the same point in space and in an energy interval
of the order of the Debye energy, which is taken to be typically of the order of $10-20\%$ of the
Fermi energy (therefore we consider off-diagonal couplings between Landau levels with $n_c-n_D
\leq n \leq n_c+n_D$, where $n_D \sim \omega_D/\omega_c$).

The excitation spectrum depends on $\Delta$ and on the dimensionality of the lattice 
\cite{5}. It
also depends on $n_c$ but appropriate rescalings yield a somewhat universal behavior for not too large
values of $\Delta$ \cite{4,5}. For very small $\Delta$ (in units of the Landau spacing) a diagonal
approximation gives good results and a gapless behavior is found at a 
series of points in the
magnetic Brillouin zone associated with the zeros in 
the real lattice (Eilenberger points) and in
other points which increase in number as $n_c$ grows. These gapless 
points contribute to a density 
of states that vanishes linearly as $\omega \rightarrow 0$. As $\Delta$ grows, 
off-diagonal terms have
to be included. Their effect is two-fold since they affect both the normal 
self-energy and the
pairing self-energy. This leads to a shift in the chemical potential 
to simulate the 3D case.
In this case it can be shown \cite{5} that the Eilenberger points 
remain gapless to all orders in the off-diagonal coupling, as long as $\Delta$ is small enough that
no band-crossings occur. For $n_c$ not too large
\begin{figure}[h]
\epsfxsize=3.4in
\epsfysize=3.4in
\epsffile{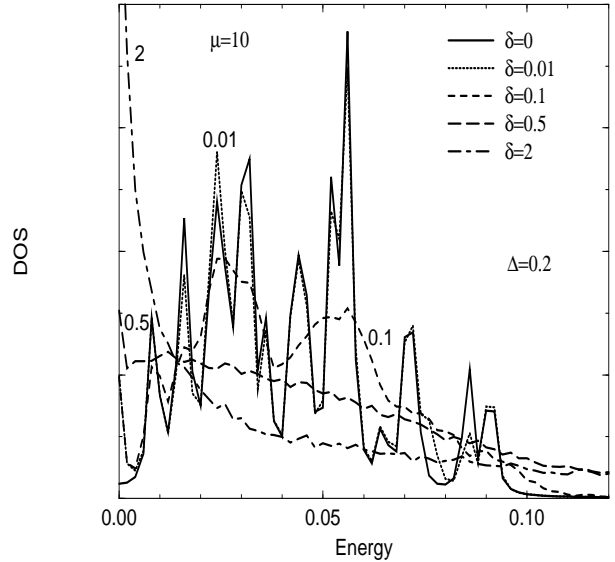}
\caption{
Density of states (DOS) (in arbitray units) as a function of energy 
for $n_c=10 \pm 2$, and $\mu=10$ for
$\Delta=0.2$ for the cases of $\delta=0,0.01,0.1,0.5,1,2$.
Only the lowest band is shown.
}
\label{fig2}
\end{figure} 
\noindent and with the inclusion of off-diagonal coupling,
a pseudo-gap opens in the spectrum since the weight of the gapless points is small (but nonzero!).
(This is peculiar to a 2d system since in this case the Fermi surface only passes through a finite
number of points, in general. In 3d there is always a value of $k_z$ such that it is possible to
find a gapless point \cite{5}). As $n_c$ grows, and for values of $\Delta$ such that no band-crossings
occur, the density of states is more and more similar to the diagonal approximation. 

We want to
see if the presence of disorder affects this gapless behavior and 
therefore we consider first the less
favorable case where a small density of states is already present in the lattice case. We therefore
consider a typical case of $n_c=10$, $n_D=2$ and $\Delta=0.2$ which was studied 
before \cite{5}. We
maintained the value of the chemical potential fixed and independent of disorder.

In the lattice case the results are very similar using either the magnetic
Brillouin formulation or the real space formulation. Due to the finiteness of the system
there is somewhat more structure in $\rho(\omega)$ but the qualitative features are the same:
a pseudogap of order $0.02$, the location of the maximum around $\omega \sim 0.07$ and
the width of order $0.1$ for the lowest band above the Fermi level. 

Considering now the case of weak disorder we performed averages over $100$ configurations for
$\delta=0,0.01,0.1,0.5,1,2$.
Our results are shown in Fig. 1.
The cases $\delta=0$ and $\delta=0.01$ are virtually indistinguishable.
As $\delta$ increases the effect of disorder is clear. For $\delta=0.1$ the width of the band is
almost unaltered but the structure is now broadened. For larger values of $\delta$ this effect is
more pronounced and the width of the band increases. In particular, the density of states for
small $\omega$ increases and the relative weight of the gapless modes increases with respect to the
maximum $\rho(\omega)$, which remains approximately at the same energy. The case of $\delta=2$ is
already very similar to the case of strong disorder where the randomness is maximized.
Even though the disorder strongly affects the $u(\vec{r})$ and $v(\vec{r})$ amplitudes
we found no evidence for localization. The lowest energy eigenvectors extend considerably
over the whole system even though they are strongly inhomogeneous.

The pseudogap of Fig. 1 found in the $2d$ case for $n_c=10$ with the chemical
potential adjusted disappears over a wide range of the order parameter if the
number of levels increases \cite{4,5}. Also, as discussed above, in $3d$ there is
always a value of $k_z$ such that the pseudogap vanishes. Keeping the chemical
potential at $\mu=10$ the pseudogap also becomes small since, even though the
Eilenberger points are not gapless in general, as long as the order parameter
is small, the gaps are very small throughout the Brillouin zone. In Fig. 2
we consider the case of $n_c=10$ and $\mu=10$ as a function of disorder. The case
$\delta=0$ shows a small finite density of states at the Fermi level (due to the
finite size considered) and a large density of states at low energies. As the
disorder increases its effect is very pronounced. The DOS broadens as before
and extends from zero energy with a zero energy value that increases as the $\delta$
increses. For strong disorder ($\delta=2$) the DOS is much larger than the
lattice result. 

These results show that in the presence of disorder the 
gapless behavior does not disappear and
is actually
enhanced. Our numerical results indicate that there is a finite density of states at zero
energy particularly in $3d$ or if the number of Landau levels is not too small.
They also confirm that the gapless behavior has a topological nature and is not 
specific to the periodic vortex lattice.

The author acknowledges helpful discussions with Zlatko Tesanovic and
partial support from PRAXIS Project /2/2.1/FIS/302/94.



\end{multicols}

\begin{references}

\bibitem{1} A.A. Abrikosov, Sov. Phys. JETP {\bf 5}, 1174 (1957).

\bibitem{2} M. Rasolt and Z. Tesanovic, Rev. Mod. Phys. {\bf 64}, 709 (1992).

\bibitem{3} S. Dukan, A.V. Andreev and Z. Tesanovic, Physica C {\bf 183}, 355 (1991).

\bibitem{4} S. Dukan and Z. Tesanovic, Phys. Rev. B {\bf 49}, 13017 (1994).

\bibitem{5} Z. Tesanovic and P.D. Sacramento, Phys. Rev. Lett. {\bf 80}, 1521 (1998).

\bibitem{6} M.R. Norman, A.H. MacDonald and H. Akera, Phys. Rev. B {\bf 51}, 5927 (1995).

\bibitem{7} C. Caroli, P.G. de Gennes and J. Matricon, Phys. Lett. {\bf 9}, 307 (1964).

\bibitem{8} P.W. Anderson, Phys. Rev. Lett. {\bf 3}, 325 (1959).

\bibitem{9} K. Ueda and T.M. Rice, in "Theory of Heavy Fermions and Valence
Fluctuations", Ed. T. Kasuya and T. Saso (Springer, Berlin, 1985).

\bibitem{10} P.A. Lee, Phys. Rev. Lett. {\bf 71}, 1887 (1993); Y. Hatsugai and P.A. Lee,
Phys. Rev. B {\bf 48}, 4204 (1993).

\bibitem{11} M. Franz, C. Kallin and A.J. Berlinsky, Phys. Rev. B {\bf 54}, R6897 (1996).

\bibitem{12} S. Haas, A.V. Balatsky, M. Sigrist and T.M. Rice, cond-mat/9703082.

\bibitem{13} A.A. Abrikosov and L.P. Gorkov, Sov. Phys. JETP {\bf 12}, 1243 (1961).

\bibitem{14} S. Dukan and Z. Tesanovic, Phys. Rev. B {\bf 56}, 838 (1997).

\bibitem{15} Z. Gedik and Z. Tesanovic, Phys. Rev. B {\bf 52}, 527 (1995).

\bibitem{16} P.G. de Gennes, Superconductivity of Metals and Alloys (Addison-Wesley, Reading, MA, 1989).

\bibitem{17} V.G. Kogan, J. Low Temp. Phys. {\bf 20}, 103 (1975).

\bibitem{18} Y.A. Bychkov and E.I. Rashba, Sov. Phys. JETP {\bf 58}, 1062 (1983).

\end{references}
\end{document}